\documentclass[5p]{elsarticle}

\usepackage{amsmath}
\usepackage{amsthm}

\usepackage{pifont}
\usepackage{ifpdf}
\usepackage{hhline}
\usepackage{color}
\usepackage[algo2e,lined,boxed,linesnumbered,ruled,vlined]{algorithm2e}
\usepackage{graphicx}
\usepackage{subfig}
\usepackage{multirow}

\newtheorem{definition}{Definition}
\newtheorem{theorem}{Theorem}
\begin{document}

\begin{frontmatter}

\title{An effective Time-Aware Map Matching process for low sampling GPS data}

\author[mymainaddress]{Paolo Cintia \corref{mycorrespondingauthor}}

\ead{paolo.cintia@isti.cnr.it}

\author[mymainaddress]{Mirco Nanni}

\address[mymainaddress]{KDD-Lab Isti CNR - Via G. Moruzzi 1, Pisa - Italy}

\begin{abstract}
In the era of the proliferation of Geo-Spatial Data, induced by
the diffusion of GPS devices, the map matching problem still 
represents an important and valuable challenge. The process of associating
a segment of the underlying road network to a GPS point gives us the chance to enrich raw data with the semantic layer provided by the roadmap, with all contextual 
information associated to it, e.g. the presence of speed limits, attraction points, changes in elevation, etc.
Most state-of-art solutions for this classical problem simply look for 
the shortest or fastest path connecting any pair of consecutive 
points in a trip.
While in some contexts that is reasonable, in this work we argue that
the shortest/fastest path assumption can be in general erroneous. 
Indeed, we show that such approaches can yield travel times 
that are significantly incoherent with the real ones, and 
propose a Time-Aware Map matching process that tries to improve 
the state-of-art by taking into account also such temporal aspect. 
Our algorithm results to be very efficient, effective on low-sampling 
data and to outperform existing solutions, as proved by experiments on  
large datasets of real GPS trajectories. 
Moreover, our algorithm is parameter-free and does not depend on specific 
characteristics of the GPS localization error and of the road network (e.g. density
of roads, road network topology, etc.).
\end{abstract}

\begin{keyword}
\texttt{Spatio-temporal Database},\texttt{Semantic enrichment},\texttt{Map Matching}

\end{keyword}

\end{frontmatter}

\section{Introduction}\label{introduction}
The widespread diffusion of location devices for personal usage, from GPS navigators to location-based services on smartphones, are making this decade the era of Geo-Spatial Data.
Coupled with the novel technologies for storing and processing large streams of data, this phenomenon is leading to the collection of massive datasets of GPS (or GPS-like) traces describing the movement of people and vehicles, as well as to the development of analysis methods and applications that use such information to extract useful knowledge.
Some examples are provided by the current studies on knowledge discovery from spatio-temporal data, based on methods like trajectory pattern mining \cite{giannotti2007trajectory} or flock mining \cite{ong2011traffic}. 
These approaches rely only on spatio-temporal features (latitude, longitude, timestamp) of raw data without considering any contextual characteristic, such as the features of road network. 

In this context, map matching, i.e. the process of associating a sequence of GPS points to a connected sequence of road segments, gives us the chance to enrich raw data with the semantic layer provided by the roadmap and all contextual 
information associated to it, e.g. the presence of speed limits, attraction points, changes in elevation, etc.

Although it is a classical and well known task in GIS literature, the map matching problem still represents an important and a valuable challenge. 
The map matching problem can be treated at two different scales, depending on the characteristics of input data, which can be made of either high-frequency or 
low-frequency samples of the real position and movement of the device. The former is mainly treated in the field of Personal Navigation Assistants, where the device is able to identify in real time the road where the user is traveling. 
The latter is common for applications dealing with smartphones or GPS-equipped black boxes installed vehicles for security or insurance purposes.
This kind of devices sample and store their location at a lower frequency to limit the battery consumption (e.g., with smartphones) or to reduce the traffic of 
data between the device and the server that stores the information. The result is a coarse-grained GPS data, harder to deal with but still with high value: this data 
represents the most reliable proxy for road network mobility.
One important issue introduced with low-frequency samples is path reconstruction. After mapping single points to the road network, 
between two consecutive points there might be a significant gap, therefore requiring strategies to reconstruct the path traversed by the vehicle or the individual.

With this work we present a significant improvement on the state-of-art of map matching for low-frequency samples, by considering two aspects that were neglected in previous literature: first, a data-driven estimation of traversal times of road segments is introduced and exploited in the evaluation of map matching alternatives; second, we perform a shift of perspective in the path reconstruction phase and remove the most common assumption adopted in literature:
 the shortest/fastest the better. (\cite{newson2009hidden},\cite{tang2012efficient},\cite{yuan2010interactive})
{\bf Inferring and exploiting segment traversal times.}
Surprisingly enough, virtually all the literature on map matching reasons in terms of length of the alternative paths, and not in terms of time requested to traverse them -- which instead is the obvious target of personal navigation systems, for instance.
Part of this phenomenon can be explained by the general lack of reliable information about travel times on road networks, which greatly compromises the applicability of traversal time-based methods in practice.
In this work, we propose to fill in the gap by exploiting the information we can infer from the same GPS data we want to match: either the instantaneous speed, where available, or estimates derived from trip length and the timestamps attached to the points.
Thus, the path reconstruction heuristics can exploit such estimates to provide an evaluation of traversal time for each alternative path.

\begin{figure}[h!]

\centering
\includegraphics[width=90mm]{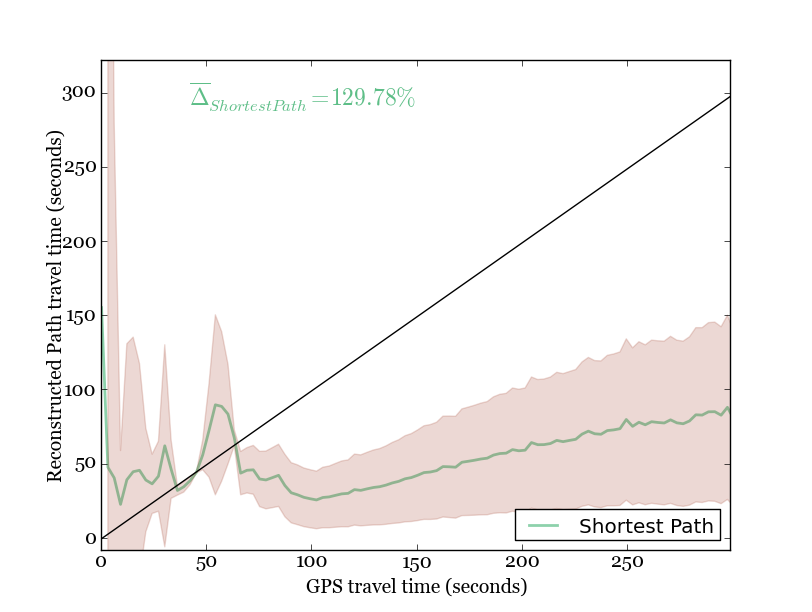}
\caption{Average travel times of reconstructed path according to the GPS travel time of the original points. The area highlight the relative standard deviation. $\overline{\Delta}$ indicates the average difference between Path and
GPS travel times.
}
\label{plot_difference}
\end{figure}

{\bf Shortest/fastest path: a questionable assumption.}
Most part of the literature on map matching assumes that the most likely path connecting two consecutive points in a trajectory is also the shortest or the fastest.
Clearly, that is inspired by the fact that real trips are just means to reach a destination B from a starting location A, without any objective other than reaching the destination in the most efficient way.
What if this seemingly obvious assumption could be violated in practice?
That would mean that, for some reason, there are trips that last longer than the minimum possible, and therefore any map matching method that looks for the shortest or fastest path would return shorter times than reality.
Since typical GPS traces also contain an accurate temporal information -- most often neglected by map matching methods -- we can actually check whether this happens or not.
Figure \ref{plot_difference} reports such an experiment; a shortest path method is applied to a real dataset of trajectories described in Section~\ref{large-dataset}, 
and the travel time according to the algorithm is compared against the real one, obtained from GPS timestamps.
It is clear that when the travel times become significant, larger than one minute, the reconstructed trips tend to be much faster than the real ones.

\begin{figure*}[t!]

\centering
\includegraphics[width=150mm,height=70mm]{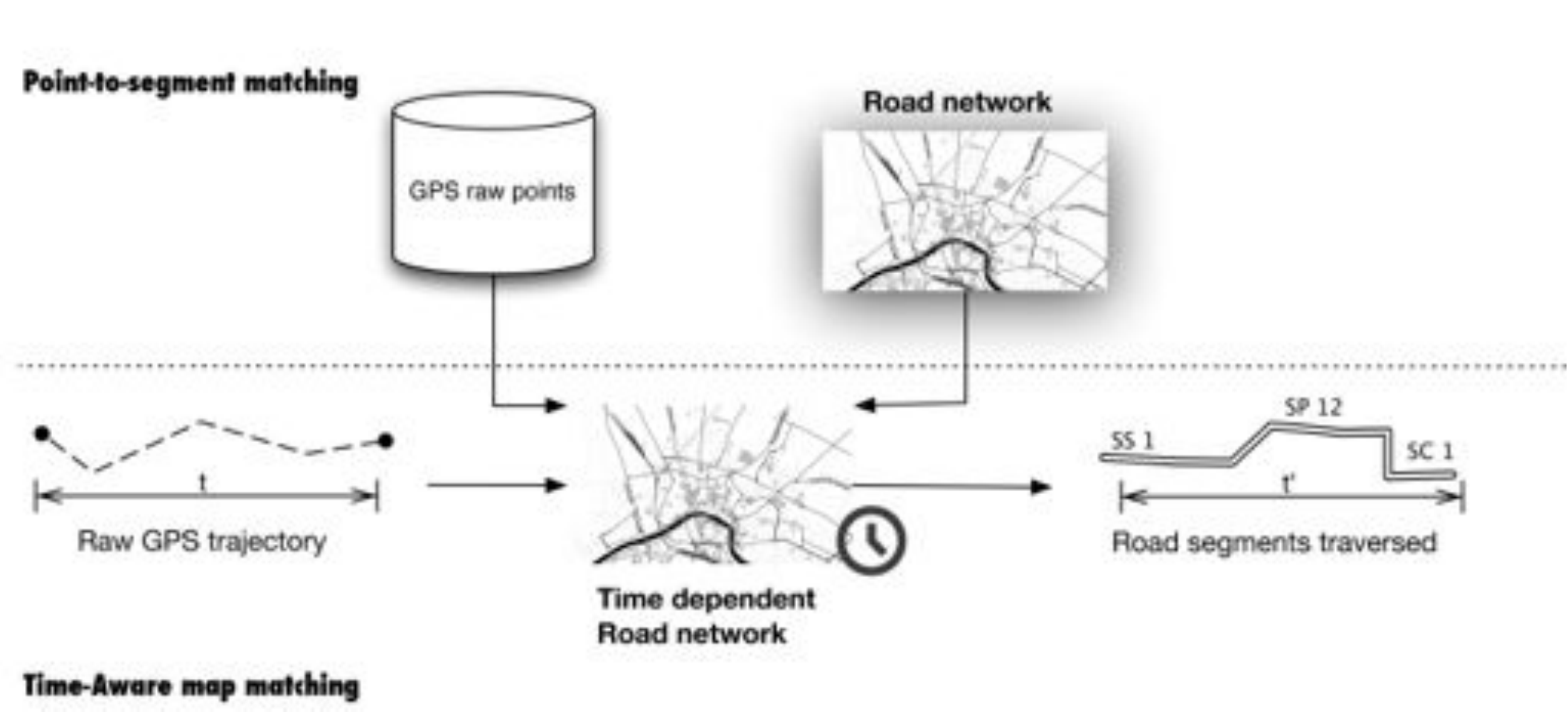}
\caption{A graphical representation of Time-Aware map-matching process }
\label{process:introduction}
\end{figure*}

We propose an effective Time-Aware Map matching process for low-sampling rate GPS data based on the reduction of temporal mismatch introduced above. Fig. \ref{process:introduction} provides
the general workflow of Time-Aware map matching. With the initial and independent point-to-matching task  we obtain a road network enriched with a precise time-dependent estimation of
travel times (see upper part of the figure), based on the method proposed in \cite{cintia2013gravity}. The core of our work is the second phase (lower part): a Time-Aware map matching algorithm that uses travel times to transform an input raw GPS trajectory -with travel time $t$- into a 
sequence of road segments with a travel time $t'\sim t$. 

\begin{figure*}[t!]

\centering
\includegraphics[width=150mm]{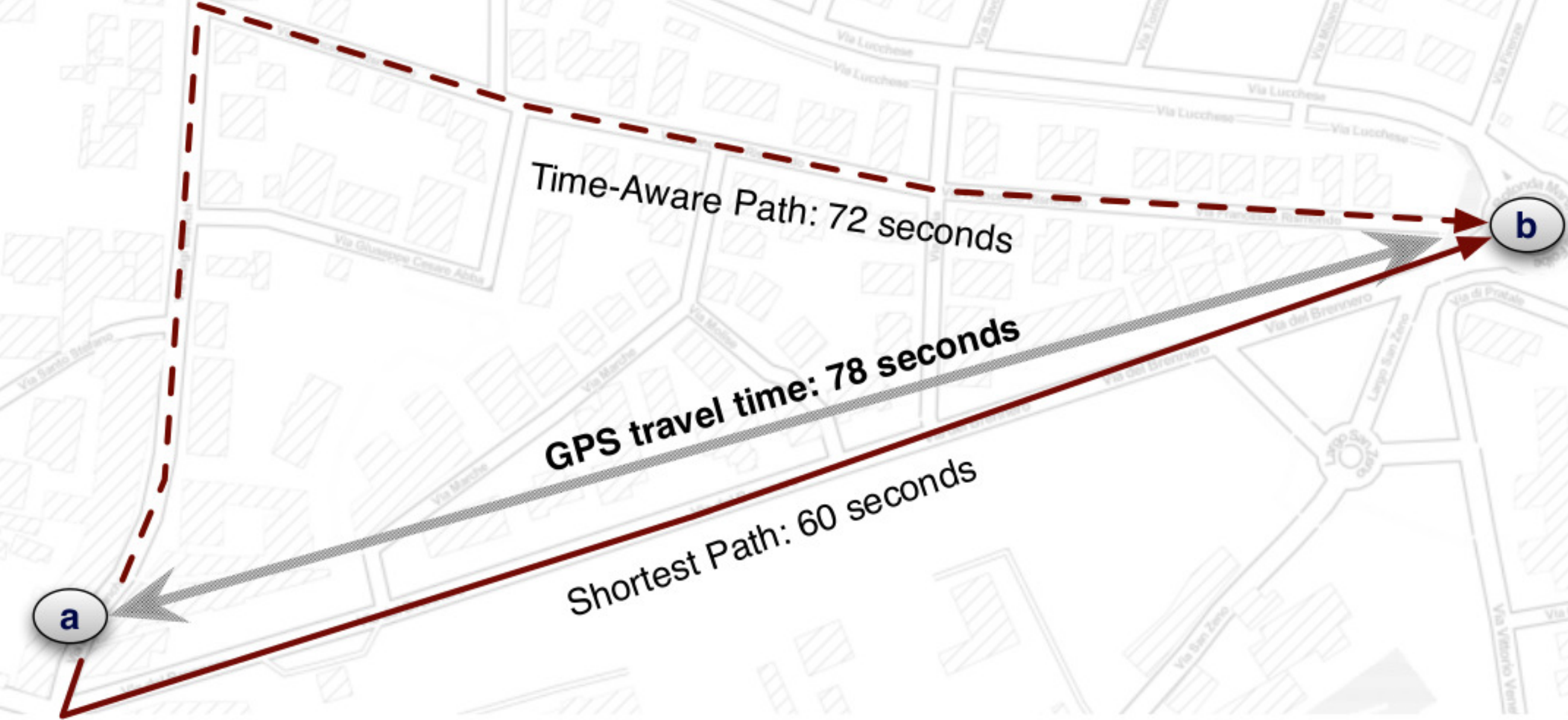}
\caption{An example from our real dataset on the problem we faced: given two GPS points $a$ and $b$ with their relative timestamp, we search for the path that mostly fit with the travel-time of input points. As depicted,
in this case the shortest path is too fast, suggesting that the user was traveling along a different path.}
\label{introduction_case}
\end{figure*}

In particular, we focus on finding the path between consecutive points that better fit 
the real travel time. 
Fig.\ref{introduction_case}
shows the idea that guided us towards the development of this new approach. The raw GPS trajectory composed by points $a$ and $b$ has a travel time of 78 seconds. Once
matched $a$ and $b$ to the corresponding road segments, thus obtaining source and destination of the path, there are two options: shortest path is also the fastest, with a travel time  60 
seconds. An alternative path, there called Time-Aware, would be more reasonable to select, since it has a more similar travel times (72 seconds). 
The main contributions of this paper can be summarized as follows:
\begin{itemize}
\item a methodology for inferring speeds and traversal times of road segments is applied, based on the principles introduced in~\cite{cintia2013gravity};
\item a novel time-aware map matching method is proposed, that takes into consideration the real traversal times as described in the raw GPS data. A proof of the complexity of method is also provided, showing its higher scalability compared with existing competitors;
\item a new methodology for evaluating the performances of map matching over large datasets, named {\em middle point test}, is introduced ad adopted;
\item a wide comparison against the state-of-art competitors is performed, based on three real datasets: a small one from SigSpatial Cup 2012 and two large ones describing, respectively, the movements of taxis in San Francisco and private vehicles in Tuscany, Italy.
\end{itemize}

The outline of the paper is as follows. Section 2 presents a survey of related works in the field of low-sampling rate map matching, while section 3 contains the definition
of our proposed algorithm. Section 4 is dedicated to the validation of the algorithm, while in section 5 all the experiments on our dataset are reported. Section 6 gives the conclusion and introduces real applications and ideas for future works.

\section{Related works}
Map matching algorithms are classified according to three categories: global, incremental and probabilistic algorithms. We will focus our
review on global and incremental algorithms, since the probabilistic approach (e.g. Kalman Filters) is used to tackle the high-sampling rate map matching problem. \\
Global algorithms solve the problem by considering the entire trajectory, the solution is obtained by searching the closest path in the map w.r.t. the input trajectory.
In \cite{alt2003matching} there is a first example of global matching algorithm: map-matching is the result of a spatial query, the resulting road network path has the
 minimum Fréchet distance w.r.t to the input trajectory. Fréchet distance is a mathematical model 
used to compare two curves: in its more common and easier illustration, 
it is viewed as the minimal length of a leash between a dog and his owner, whom are walking on different curves. The complexity of this approach is quite high: $O(nm\log^2nm)$, with $n$ as the number of trajectory points and $m$ the number of road network edges.
In \cite{chen2011approximate} a more efficient version of Fréchet distance computing algorithm has been provided. \\
The main issue for global algorithms is the purely geometric approach. All the characteristics of the road network are ignored, the matching is only based on the research
 of a similar curve. It is obvious to notice that there will never be a low-sampled trajectory completely equal to a path in the network: this means that there is not a precise definition of the optimum to reach. 
In \cite{brakatsoulas2005map} Fréchet distance is even used for a quality evaluation of the results obtained.
The incremental approach for low-sampling map matching is based on joining optimal local searches. The local optimum is represented by the most probable path between two 
consecutive matched GPS points. 
IVMM algorithm (\cite{yuan2010interactive}) is one of the state-of-the-art incremental map matching algorithm we used to compare the results of our work. The matching 
process is done through consecutive steps; first of all, a preliminary refinement is done by dropping vertex according 
to a spatial range query. Then, the matching probability is computed assuming the GPS error with a normal distribution; this position probability is combined with a 
transition probability, that is the ratio between the 
euclidean distance of two candidates and their shortest path distance on the road network. Furthermore, a temporal analysis is also considered: the cosine similarity 
between the travel-time (according to road speed limits) of 
the shortest path analyzed and the real time difference between the two GPS points. The last step of IVMM is a mutual influence modeling, used to decide the path between 
each consecutive points by considering also at the global trajectory. 
In this approach, there are several not verified assumptions: first of all a driver should always follow the shortest path. Moreover,the radius of the range query is 
arbitrary and the GPS error is assumed as Gaussian,
with fixed parameters. Furthermore, the travel-time of road edges is obtained according to road speed constraints. These constraints, especially on a city road network, 
could be considered as arbitrary. In \cite{cintia2013gravity} a method to get rid of this assumption has been proposed: a gravity model is used to associate a GPS point, with his speed, to a road segment, 
choosing between the k-nearest neighbor. In the following sections we are giving an 
exhaustive description of this method. \\
A particular case of probabilistic approach used to deal with low-sampling rate data is \cite{newson2009hidden}, where authors developed a map matching algorithm based on 
the well known Hidden Markov Model. This algorithm has a weak point on its highly dependance on two parameters, both of them obtained from the ground truth, i.e. from the 
road segments actually traversed from the vehicle; this kind of data are not available in a real application scenario of a map matching algorithm. Another factor of weakness 
is its complexity: for each trajectory, Viterbi algorithm takes $O(|C| * |S|^{2})  $ to find a solution, where $C$ is the set of transitions between segments and $S$ the
set of all the segments candidates to be matched with a point of the input trajectory. It is worth to point out that for all the transitions in $C$ the shortest path is
 computed, thus adding the complexity of this computation, that is $\sum\limits_{p \in P} |C_{p}|^{2}  *(|E| + |V| log |V|)$. Here $C_{p}$ is the subset of candidates 
segments associable to the single GPS point $p$. We compared our algorithm to \cite{newson2009hidden} - only on the dataset where a ground truth was provided - showing how it is overperfoming in terms of accuracy. 
Given the limits of this approach, it has not been possible to test this algorithm on our real GPS dataset. 

As mentioned before, we used the dataset of map matching challenge ACM SigSpatial cup 2012 to validate the results of our work. Thus, we also chose the winner of the contest (\cite{tang2012efficient}) as a competitor for our algorithm; 
even though that algorithm is designed for high-sampling data, authors states it is also supposed to work with low-sampling data. This algorithm relies completely on the topological characteristics of the network: 
candidate segments to be matched are chosen with successive filtering. The first filter is spatial, all the vertex farther than 18 meters from the GPS points are dropped. Then, all the vertex with an angle difference w.r.t. 
point direction greater than 90 degree are excluded. The remaining vertex are selected as candidates if they are on the same edge or they are connected by a path on the road network. Once the candidates has been chosen, every pair of
consecutive candidates belonging to the shortest path are matched. This method is fast and reliable with high sampled data, while it is less accurate with low sampling data. \\
Another interesting approach we did not use for a comparison is represented by \cite{li2013large}: a data-driven map-matching method where the knowledge from a large scale trajectories dataset is used to 
enrich the classic map matching approach. It represents a new approach, but it cannot be applied on the small data set we used 
for testing.

\section{Time-Aware Map Matching}

A map matching process is based on two main steps: a point-to-segment matching process and an heuristic process to choose the path between the possible candidates. These tasks are 
modeled according to the data type the map matching is designed for. The two steps are well separated, as highlighted in Fig. \ref{process:introduction}: first of all, we match the gps points 
to a road segment, 
then we reconstruct the path between every two consecutive matched segments with a Time-Aware heuristic.

All the state-of-the-art methods rely on what we called the ``shortest path'' assumption. As described before, in literature there are different approaches to map matching founded on a large variety of heuristics. The common point for all of
them is the use of the shortest path to connect two consecutive matched GPS points. The underlying assumption is: a driver is always choosing the shortest path. This strong assumption could not be true. We released 
this assumption by introducing our Time-Aware heuristic: the path between two matched GPS points is the most compatible with the real travel time. 
In Fig. \ref{plot_difference} the difference between real GPS time and path time of shortest-path reconstructed trajectories is highlighted. 
The curve represents the average shortest path time $y$ for each pair of consecutive GPS points with GPS time $x$. The line indicates the optimal case, when the shortest path travel time is equal to real GPS time.
The error bars provide an estimation of the error that affect this heuristic. For reading purpose, error bars are related to aggregated bins. As showed, for our 1M+ trajectories dataset the average difference between real time and shortest path time  is more than 120\%. 
The goal of our work is to develop a new map-matching algorithm able to minimize the difference between selected path travel time and real time, then we show how this minimization improves the results in terms of accuracy.

\subsection{Point-to-segment matching}

The first phase of our map-matching algorithm considers each point separately, matching it to a segment in the road network.

\begin{definition}
Given a location point $p$ and a road network $N=(V,E)$, with $V$ the set of vertices and $E$ the set of segments, 
we define the point-to-segment matching of point $p$ as the process of associating $p$ to a segment $e \in E$. 
\end{definition}

The main issue of point-to-segment matching is represented by the localization error of each point. Such error is variable from few meters to tens of meters, 
depending on weather conditions and satellites visibility. 
The state-of-the art map-matching methods assume that the localization error follows a Gaussian distribution, with fixed mean and standard deviation. This is a really strong assumption 
that could affect the results: 
on our dataset we saw that error varies depending on different factors, e.g. in the city center, the presence of big buildings affects the precision of localization. 
In order to prevent any assumption of that kind, in this work we adopt the gravity model proposed in \cite{cintia2013gravity}.
In such model, a point $p$ is assigned to the segment $e \in E$ that {\em attracts} it most. Following there is the definition of attraction computation and speed estimation according
to the distance and the heading difference between $p$ and $e$. 
This implies that each point has to be assigned to a direction, either provided as input data or derived from it.

A further exploitation of the point-to-segment matching is that, where available, any contextual information associated to the each point can be transferred to the corresponding road segments, 
thus enriching the existing background knowledge on the road network. The basic example, also exploited in this work, is that of the instantaneous speed of vehicles for each point, 
which enables to estimate real speeds (and therefore travel times) over the road segments, possibly dependent on the time of the day. In our work, the point-matching have been
also applied to estimate travel times for each segment of the road network. To the best of our knowledge, this is the more accurate
method to estimate the typical speed of road segments.  \\

\label{gravitymodel}
Let $O$ be the set of GPS points,$R$  and the set of road segments,  we define the \emph{attraction} of a segment $j$ for a point $i$ as:
$w_{(o_{i},e{j})} = w^{d}_{(o_{i},e{j})} \cdot w^{\theta}_{(o_{i},e{j})}$ where 
\begin{itemize}
\item $w_{(o_{i},e_{j})}^{d} = 1 - \frac{dist(o_{i}, e_{j})}{\sum\limits_{e_{k} \in E} dist(o_{i}, e_{k})}$; 
\item $w_{(o_{i},e_{j})}^{\theta} = 1 - \frac{ang(o_{i}, e_{j})}{\sum\limits_{e_{k} \in E} ang(o_{i}, e_{k})}$;  
\end{itemize} 
Here $dist$ is the Euclidean distance between a point and a segment while $ang$ is the relative direction difference. The direction is measured in degrees,
where 0 degrees indicates north direction while 180 degrees reprensent south direction. For every GPS point of our bigger dataset, this value
comes directly from GPS device.
Therefore the force of attraction of a segment over a point is defined by the combination of these two dimensions as:
\begin{definition}
Let $o_{i} \in O$ be a Gps point. The gravity force function between $o_{i}$ and a segments $e_{j} \in E$ is  $GF(o_{i},e_{j}) = w_{(o_{i},e_{j})}^{d} \times w_{(o_{i},e_{j})}^{\theta}$.
$o_{i}$ is then assigned to the segment with the most powerful force: $\sigma(o_{i}, E) = argmax_{e_{j} \in R}(GF(o_{i},e_{j}))$.
\end{definition}

\begin{figure}[!t]
\begin{center}
\includegraphics[width=.40\textwidth]{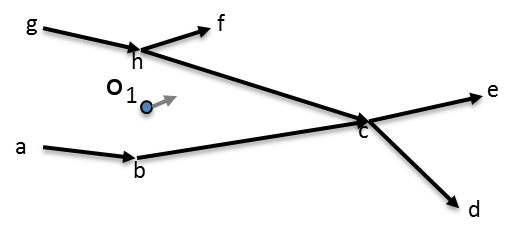}
\includegraphics[width=.50\textwidth]{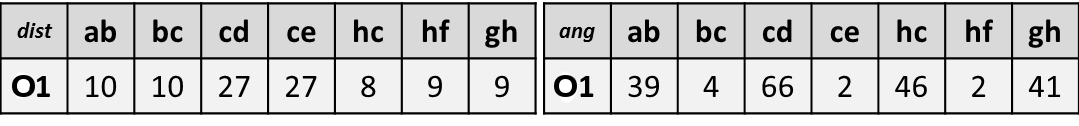}
\caption{An example of GPS point and candidate segments used to explain the gravity model (Top) and the two distances: Euclidean distance (right) and Angular (left) 
between each point-segment pair.}\label{fig:example}
\end{center}
\end{figure}

To better understand the idea under the definitions, we present below a complete example (Figure  \ref{fig:example}) of how forces are computed and how 
points are assigned to segments.
In Figure \ref{fig:example} an example of points and segments is shown, points are attracted by all the segments with different forces and suddenly they fall over one of them.
For example the point $O_{1}$ undergoes the following forces:

\begin{math}
\\
GF_{(o_{1},r_{ab})}^{d} = (\frac{1-10}{100})(\frac{1-39}{200}) =0.2745 \\
GF_{(o_{1},r_{bc})}^{d} = (\frac{1-10}{100})(\frac{1-4}{200}) =0.882 \\
GF_{(o_{1},r_{cd})}^{d} = (\frac{1-28}{100})(\frac{1-66}{200}) =0.4154\\
GF_{(o_{1},r_{ce})}^{d} = (\frac{1-28}{100})(\frac{1-2}{200}) =0.6138\\
GF_{(o_{1},r_{hc})}^{d} = (\frac{1-8}{100})(\frac{1-46}{200}) =0.7084\\
GF_{(o_{1},r_{hf})}^{d} = (\frac{1-9}{100})(\frac{1-2}{200}) =\textbf{0.9108}\\
GF_{(o_{1},r_{gh})}^{d} = (\frac{1-9}{100})(\frac{1-41}{200}) =0.78705\\
\end{math}

Therefore the point is attracted mostly by the segment $hf$ or more formally $\sigma(o_{i}, R) = r_{hf}$. 

To speed up the calculation, the candidate segments are the $k$ nearest segments for $p$. 
Results in \cite{cintia2013gravity} show that $k=8$ is a good compromise.

\subsubsection{Travel times estimation}
In order to obtain reliable travel times for each road segment, we define a travel time function based on all the GPS points of our dataset. The function described below
considers the segment matched for all the GPS points, with the relative weights, to compute the typical segment speed by calculing a weighted mean. 

\begin{definition}
Given a set of GPS points $O =\{o_{1} \ldots o_{m}\}$ where $o_{i} = (p_{i},d_{i},s_{i})$ has its spatial position $p_{i}$, its direction $d_{i}$ and its speed $s_{i}$, 
a set of road segments $E = \{e_{1} \ldots e_{n}\}$ and having an  function $\sigma(o_{i},E) = (w_{(o_{i},e{j})},e_{j})$ assigning the point $o_{i}$ to 
the segment $e_{j} \in E$ with a confidence value $w_{(o_{i},e{j})}$, it is possible to estimate the speed over the segment as:
$$Speed(O,R,e_{j}) =  \frac{\sum\limits_{o_{i} \in O_{j}}w_{(o_{i},e{j})} \cdot o_{i}.s}{\sum\limits_{o_{i} \in O_{j}}w_{(o_{i},e{j})}}$$		
Where $O_{j}=\{o_{i} | \sigma(o_{i},E)=(w_{(o_{i},e{j})},e_{j})\}$.
\end{definition}



%

\subsection{Path reconstruction heuristics}

The state-of-art methods for map matching on low sampling rate data rely on the ``shortest-path'' assumption: the most likely path between two consecutive matched points is always the shortest one. 
That clearly stems from various assumptions, such as the fact that the trip performed has the unique objective of reaching the destination, without any other goal -- e.g., traversing more pleasant roads, performing quick bring-and-get tasks, avoiding crowded roads (with or without traffic jams) -- and/or the fact that drivers really know what is the most efficient path. 
In our work we release these assumptions, aiming to realize a more realistic heuristic. \\
Exploiting the results in \cite{cintia2013gravity}, each road segment is associated with its estimated travel time, with an error of $10.4\%$. 
In the following, we leverage this information to built a time-aware heuristics that associates each pair of points in a user's trip to the path that best fits with its travel time. \\

The problem of finding a set of road segments whose travel time fit with the travel time of two GPS points remind the knapsack problem, which is NP-hard and solvable with a linear programming approach. 
Yet, in our case we have additional constraints and requirements. First of all, the chosen road segments have to be connected to each other, in order to form a path in the road network. 
Furthermore, the travel time compatibility requirement alone might be not sufficient to obtain reasonable path reconstructions. For instance, if the real path is slower than expected (w.r.t. estimated travel times over segments), using only the travel time difference as optimization criterion
could lead to odd results: in order to reach the optimum, wrong segments might be added, only to create artificial detours that increase the overall travel times to better fit the real one. 
In order to take care of such extreme behaviours as well as to provide computationally sustainable solutions, we propose an heuristic based on a routing approach: 
indeed, routing algorithms take implicitly account of road network topology, and their complexity is lower than linear programming approaches.\\

\begin{figure*}[t!]

\centering
\includegraphics[width=160mm,height=70mm]{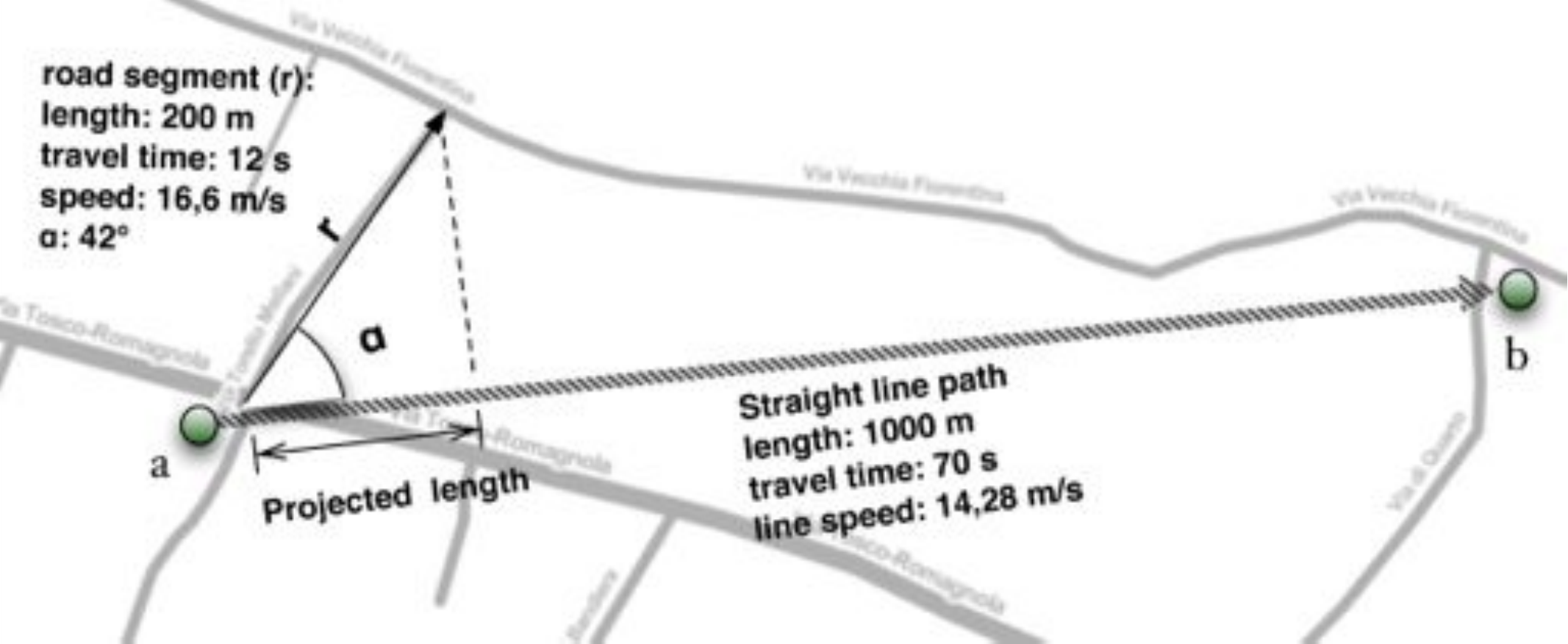}
\caption{The computation of timecost for segment $r$: the length of segment is projected onto the straight line trajectory between GPS points $a$ and $b$, 
then the timecost is computed as the difference between the length and the supposed length of the segment, according to the straight line speed of the car.}
\label{timecost}
\end{figure*}

The method we propose is based on Dijkstra's shortest path algorithm, using a time-aware heuristic to evaluate the cost of every road segment according on how it fits with the trajectory's real travel time.
The core of our approach is the introduction of a $Timecost$ function: given a source and a target node, Dijkstra algorithm uses Timecost to evaluate the cost of every segment 
examined to find the shortest path. 
Hence, we obtain a path that is the shortest in terms of timecost. We can define the solution we found as a path with two constraints: (i) acyclicity and (ii) highest similarity
line speed w.r.t. the real line speed. The acyclicity of the solution is guaranteed by Dijkstra algorithm, while the similarity with the real line speed is the result of $Timecost$ minimization. In other words, the solution found is the
path with the more similar travel time w.r.t. to the real GPS travel time according to network and speed constraints.
Below we provide the details about how Timecost is computed and prove how the choice of the less expensive path according to Timecost finds a solution for the problem we tackled,
yet satisfying the constraints we introduced. More formally, we prove that the minimization of
$timecost$ yield the more similar path in the road network according to real GPS line speed. As a result, path travel time will approximate the real travel time, yet 
respecting network costraints. \\
In fig. \ref{timecost} there is a graphical representation of Timecost computation, while in fig. \ref{timecost-pseudo} Timecost is analitically showed through pseudocode; input $linespeed$ is defined as the ratio between 
the distance between the two consecutive GPS points $a$ and $b$ and the related travel time, i.e. the time difference between $b$ and $a$ timestamps. $heading$ indicates the angle of the straight line between $a$ and $b$ w.r.t. north heading.
The timecost of a segment is then obtained by projecting its length onto the straight line defined by $a$ and $b$.
Expected length ($exp\_length$) is the length of the segment if its projected speed would be equal to the real
line speed of the car.
Projected speed defines the relation between the typical speed of a segment and its projection onto the straight line between $a$ and $b$. 

\begin{definition}
We define the \emph{timecost} for a road segment $r$ as the difference between the length of $r$ and the expected length of $r$ according to the straight line speed of the vehicle.
\end{definition}

\begin{theorem}

Let $p$ be a path in the road network connecting two nodes $a$ and $b$ such that $Tc(p) = \arg\min_{p \in P } Tc(p)$, with $Tc$ as the timecost function and $P$ the set of all the possible paths connecting $a$ and $b$. For
each $p' \neq p$ we have:
\begin{displaymath} 
Tc(p)<Tc(p') \implies |r.speed-p.lspeed| \leq |r.speed-p'.lspeed|
\end{displaymath}
where $r.speed$ is the ratio between the straight line distance between $a$ and $b$ and the travel time recorded by GPS device (car linespeed)
 and $p.lspeed$ is the ratio between the straight line distance between $a$ and $b$ and $p.traveltime$ (path linespeed).
\end{theorem}

\begin{proof}
Suppose on the contrary there exists a path $p''$ such that
\begin{displaymath}
Tc(p)<Tc(p'') \implies |r.speed-p.lspeed| > |r.speed-p''.lspeed|.
\end{displaymath}
For the sake of simplicity, we will assume $p''$ identical to $p$ except for the total travel time, and, accordingly, their $lspeed$ . Then, for the definition of Timecost function (fig. \ref{timecost}),
we have
\begin{gather*}
|p.length_{expected} -p.length | < |p''.length_{expected}-p''.length | 
\\ \implies |r.speed-p.lspeed| > |r.speed-p''.lspeed|
\end{gather*}
Since $p.length=p''.length$,  we can simplify:
\begin{displaymath}
p.length_{expected}  < p''.length_{expected} \implies p.lspeed > p''.lspeed
\end{displaymath}
Recalling the definition of $p.lspeed$  we have then:
\begin{gather*}
p.length_{expected}  < p''.length_{expected}
\\ \implies \frac{1}{p.traveltime} > \frac{1}{p''.traveltime}
\end{gather*}
\newline
Since $length_{expected}$ is defined, for a segment $e$, as 
\newline
\begin{math}
\frac{length(e) * time_{expected}}{time(e)}
\end{math} 
(see Fig. \ref{timecost-pseudo}) and given our initial assumptions on $p$ and $p''$, the expression become:

\begin{math}
\centering
p.traveltime < p''.traveltime 
 \implies p.traveltime > p''.traveltime
\end{math}\\
that is impossible.

\end{proof}

 \begin{algorithm2e} [h]
\scriptsize{
\KwIn{A road segment $e$; a travel heading $h$; a linespeed $ls$;}
\KwOut{Timecost for segment $e$}
\BlankLine
$\alpha= | heading(e)-h |$;\\
$length_{projected}=length(e)* cos(\alpha)$;\\
$time_{expected}=\frac{length_{projected}}{ls}$;\\
$length_{expected}=\frac{length(e) * time_{expected} }{time(e)}$;\\
\Return {$|length_{expected}-length(e)|$};

\caption{$Pseudocode for Timecost computing$}
\label{timecost-pseudo}}
\end{algorithm2e}

In figure \ref{timecost} there is a graphic sample about the computation of timecost for a segment, with the definitions of \textit{projected length,expected time, projected speed and expected length}.
As showed, the more a road segment is compatible with the line speed of the car, the lower will be its timecost.  
Our Time Aware Dijkstra algorithm is illustrated in figure \ref{time-aware-dijkstra}. The main difference with the classic implementation is the evaluation of the weight of the edges  
the remaining real time is used to compute the straight line speed of the car. It is worth to notice how the line speed to fit the reconstructed path with changes according to already computed path. For each edge analyzed, $timeleft$ is obtained
as the difference between of the path traveltime until the parent node and the real traveltime. This allows us to construct a solution that best fit the real line speed at each step. A particular case arise when $timeleft<0$, that is the case
when the car is traveling at higher speed than road network edges usual speed. In this situation, the algorithm uses the global linespeed as goal linespeed to fit the remaining path. This still satisfies the constraints of the solution we are 
searching for. 
More formally, if $timeleft<0$ timecost for the edge is calculated using the straight line from the node currently examined and the remaining time to evaluate the current straight line speed. Otherwise, the the source-target 
straight line speed is used.

 \begin{algorithm2e} [h]
\scriptsize{
\KwIn{A road graph $G$; a source node $s$; a target node $t$;a travel time $t$}
\KwOut{A list of traversed road segments $id$.}
\BlankLine
$Q=s$;\\
$parent[s]=0$;\\
$weight[s]=0$;\\
\While {$Q \neq \phi$} {
        $n=Q.pop()$;\\
        \If {$n \neq t$} {

        \For{$edge=(n,v) \in neighbors(n)$}{
	      $timeleft=t-path_time(s,v)$;\\
              $heading=edge.heading;$\\
              \If {$timeleft<0$}{
                  $line=euclidean_distance(n,t)$;\\
                  $linespeed=\frac{line}{timeleft}$;\\
              \Else{
                  $line=euclidean_distance(s,t)$;\\
                  $linespeed=\frac{line}{traveltime}$;\\
                   }
              }
              $w=Timecost(edge,heading,linespeed)$;\\
              \If {$weight[v]>weight[n]+w$}{
                  $parent[v]=n$;\\
                  $weight[v]=weight[n]+w$;\\
                  }
              $Q.push(v)$
	}
	}
        \Else{
             
             $p=parent[t]$
             \While{$p \neq 0$}
                   {$path.append(edge=(parent[p],p)$;\\
                    $p=parent[p]$;\\}
             \Return {$reverse(path)$}; 
        }
}

\caption{$Pseudocode for Time-aware Dijkstra algorithm$}
\label{time-aware-dijkstra}}
\end{algorithm2e}
 
\subsection{Analysis of the algorithm proposed}

The map matching algorithm we propose is split into two task: point-matching and time aware shortest path. The complexity of the former is due to the search of the k-nearest segments. 
In our implementation, we used a Generalized Search Tree to index the geometries of the road network: the complexity is then $O(log |E|)$ with $E$ as the set of road segments.
A further refinement is possible: a bounding box based on latitude and longitude of the point to be matched can decrease the set of segments in which to search. The computation of the travel time of each segment is a once running process, and it runs in $O(log|E|) * |P|)$, with $P$ as the number of GPS points.\\
The complexity of Time-aware Dijkstra depends from the number of nodes needed to visit until finding the target node. Since the road network is a really sparse graph (most of the nodes have only one neighbor), 
the visited nodes are mainly depending by the distance between the source node and the target node.  According to the characteristics of our dataset (see table \ref{dataset}), the average distance between two 
consecutive GPS points is $1,500 m$. Dividing the space with a grid of 1.5 km square cells, the average number of segments for every cell is  $65$. Since the complexity of Dijkstra's algorithm
is $O(|E| + |V| log |V|)$, with $E$  as the number of edges and $|V|$ as the number of nodes.
Furthermore, Time-Aware map matching could be run in parallel. For every two consecutive points of a trajectory, the problems of finding the respective paths are independent from each other. This represents an 
important improvement w.r.t. global algorithms. 

\subsection{Implementation details}
The Time-Aware map matching algorithm is based, as stated, on Dijkstra's algorithm for shortest path. We chose this algorithm instead of other faster options because of some important properties. A* algorithm is an heuristic algorithm
to find shortest path: thanks to the heuristic approach the algorithm can limit the number of visited nodes, so obtaining a lower complexity w.r.t. to Dijkstra ($O(|E|)$). However, to be optimal A* requires a heuristic function that does not overestimate the distance to the target node. This makes
the use of A* algorithm with our Time-Aware approach impossible, since we are not able to define that function using our Timecost instead of the Euclidean distance as it is in usual implementations of A*. In other words, Timecost is already
an estimation and it also changes depending on the characteristics of the road network: to use it for an heuristic function for A* we should be able to estimate, in terms of Timecost, the distance between two nodes. 
By definition of Timecost, 
this value changes depending on some real parameters, such as car line speed. Then, in our Time-Aware scenario Dijkstra's algorithm ensures reliable results, since it is not 
depending on any heuristics, returning the shortest path after 
having visited all the nodes. Another particular detail is related to the point-matching task. As explained, this task is performed through a gravity model that returns the 
most probable segment to match the input GPS point, without any indication
about the exact point of the road segment where the GPS point should fall. Since this information is useful, we used the approach introduced in \cite{giovannini2011novel}: the 
input GPS point belongs to the closer point of the matched road
segment, we are then able to create a new node on the road network connecting the matched point and the target of the matched segment, preserving all the characteristic of the matched segment. 
Of course, travel time is modified according to the new length of the segment. This procedure guarantees a still better estimation of Timecost, resulting in a better global accuracy for our Time-Aware algorithm. \\
For a better evaluation of our work, we developed a QGIS plugin based on our Time-Aware algorithm. QGIS is  is a cross-platform free and open-source desktop geographic information systems (GIS) application 
that provides data viewing, editing, and analysis capabilities. The plugin allows to match a trajectories layer to a road network layer, obtaining the map-matched
data related to input trajectories. An alpha version with some sample shapefiles is available at: http://kdd.isti.cnr.it/software/time-aware-matching-plugin-qgis. We intend to go on developing the plugin, with the goal
of a next public release.  

\section{Experimental results}

The evaluation of our algorithm has been done through various tests. First of all, we compared the results of Time-Aware algorithm applied to the ACM Sigspatial cup dataset, that is
a GPS dataset with correct traversed roads reported by drivers. This dataset gave us the possibility to directly assess the accuracy of Time-Aware algorithm and the competitors. Then, 
we perfomed other two tests on two large datasets, OctoPisa and San Francisco cabs (see datasets characteristics on tab. \ref{dataset}): using data from OctoPisa we checked the coherence of our
Time-aware map matched trajectories w.r.t. Variable Message Panel data, that is the number of cars passing by for some particular roads; on both datasets we 
defined a ``middle-point test'' to check Time-Aware accuracy time with half-sampled GPS trajectories. Furthermore, we showed how Time-Aware is better than competitors on fitting
the original GPS travel time (see Fig. \ref{time_difference_Compare}).

\begin{figure*}[h!]
\begin{minipage}[b]{0.45\linewidth}
\includegraphics[width=90mm]{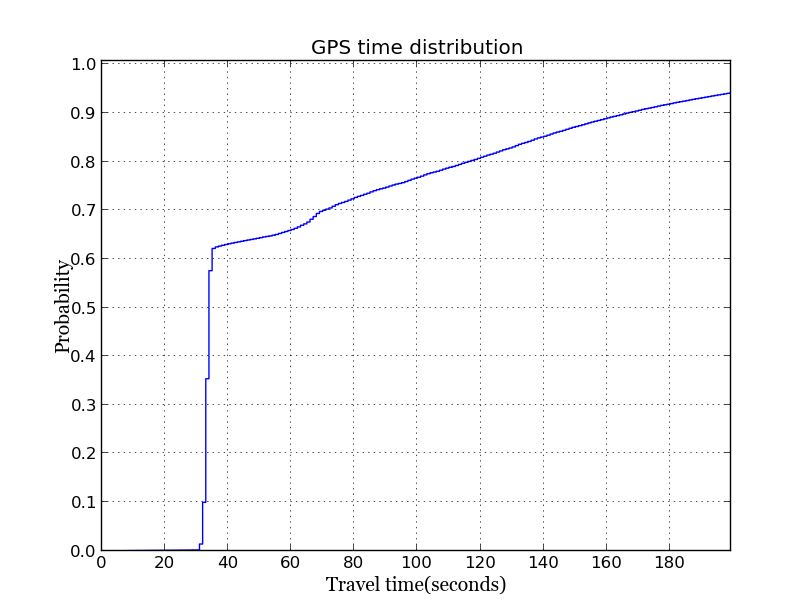}
\label{octopisa-cumulative}

\end{minipage}
\quad
\begin{minipage}[b]{0.45\linewidth}
\includegraphics[width=90mm]{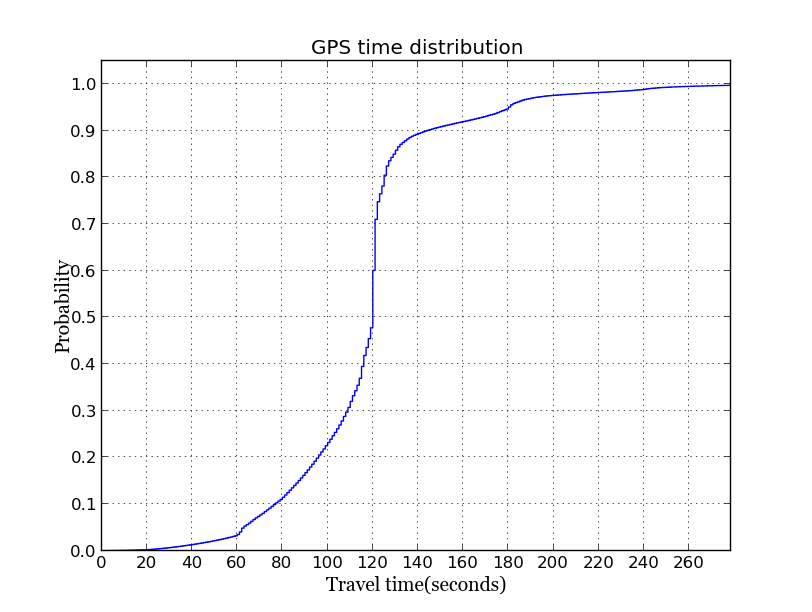}
 \label{sanfrancisco-cumulative}

\end{minipage}

\caption{GPS sampling rate cumulative distribution for Octopisa (left) and San Francisco cabs (right) dataset}
\label{sampling_Distribution}
\end{figure*}
\subsection{ACM SigSpatial Cup dataset}

Before the application of our Time-Aware map matching to the real dataset, we evaluated its accuracy. In order to do this, we relied on the dataset 
of ACM SigSpatial cup 2012 (\cite{ali2012acm}): 10 trajectories with the correct route annotated by drivers. We compared our algorithm with two 
competitors: Tang-Zhu-Xiao algorithm (\cite{tang2012efficient}) and IVMM (\cite{yuan2010interactive}). The former is the winner of the SigSpatial Cup, 
it is designed for high sampling rate data, but authors state it is  performing well with low sampling data too. The latter is the state-of-the-art 
map-matching method for low sampling rate. \\
The SigSpatial Cup dataset trajectories have a sampling rate of 1Hz: this gave us the possibility to check the performance of our method and the competitors using different sampling rates
by selecting different lower-sampled subtrajectories extracted from the original ones. Besides our Time-Aware map-matching algorithm, 
we also developed a simpler version, using the same Gravity Model (\cite{cintia2013gravity}) for the point-matching task and a Shortest Path heuristics.
The accuracy of Gravity Model matching for the validation dataset is reported on Tab. \ref{gravity-accuracy}. It is simply computed by comparing the number of correct assignments over
the total number of points.
 Since we did not have a proper dataset to compute
the travel time of all the segments of the road network, we derived these values from the trajectories provided; the travel time of segments without any GPS point 
associated has been assigned according to the typical speed recorded for its same-class neighbors.  We started this spreading from the neighbors of 
the segments with at least one gps point associated. \\
The metric used to evaluate the correctness of the map matching algorithms is the same used in the SigSpatial Cup, 
defined as the ratio between segments correctly matched and the total number of correct segments: 
\begin{center}
 
\begin{math}
Accuracy= \frac{|Correct\ segments\ matched|} {|Ground\ truth\ Segments|}
\end{math}

\end{center}

\begin{center}
\begin{table}[h!]
\begin{tabular}{|l|l|l|}
\hline
\textbf{Sampling} & \textbf{Points n.} & \textbf{Accuracy} \\ \hline
1s                & 13345              & 0.9732    \\ \hline
10s               & 1352               & 0.9415    \\ \hline
30s               & 462                & 0.9545    \\ \hline
90s               & 165                & 0.9939   \\ \hline
120s              & 129                & 0.9689    \\ \hline
\end{tabular}
\caption{Gravity Model point-matching accuracy}
\label{gravity-accuracy}
\end{table}
\end{center}

\subsubsection{Accuracy results}
\begin{figure}[h!]

\centering
\hspace{-2cm}\includegraphics[width=95mm,height=60mm]{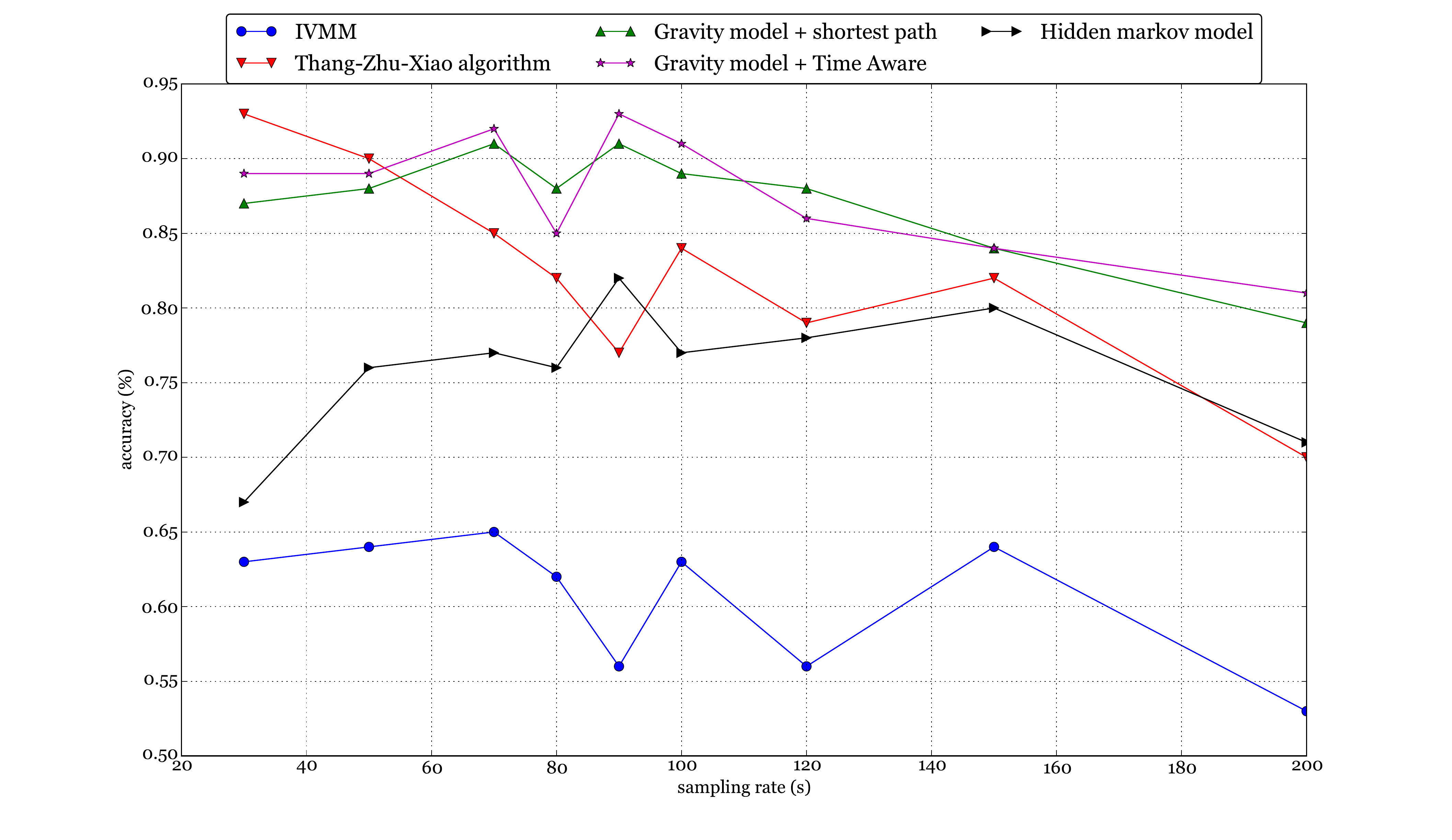}
\caption{Precision of Time-Aware map matching w.r.t. competitors}
\label{precision}
\end{figure}

The comparison between our Time-Aware map matching and the competitors is shown in Figure \ref{precision}. Our algorithm is outperforming IVMM and Tang-Zhu-Xiao algorithm. 
Furthermore, Time-Aware heuristic is working better than the simple shortest path, confirming our starting idea: 
assuming all the drivers traveling along shortest path lead to slightly inaccurate
results. 
\\
Another difference between Time-Aware map matching and the competitors is the use of parameters. The only parameter used by Time-Aware map matching is 
the number of k-nearest neighbor in the point-segment matching process; as stated in \cite{cintia2013gravity}, using $k=8$ is a good choice, since the accuracy is not increasing 
using greater values for $k$. IVMM is using a range query with radius $\epsilon = 100m$ to choose the initial candidates for the matching with a GPS fix.  
Then, a Gaussian distribution with $\mu = 15m$ and $\chi = 10m$ is used to evaluate the every candidate. Tang-Zhu-Xiao algorithm is relying on a set of parameters as well. 
The initial choice is made by selecting the 50-closest mini-vertex of the road network. Then, all the candidates at distance greater than $18m$ from the GPS fix are discarded.
In the following evaluation of the candidates, some scaling factors adopted from \cite{brakatsoulas2005map} are used to compute the score for a match:  $\mu_{\alpha}=10$, $c_{\alpha}=4$, $\mu_{d}=0.17$ and $c_
{d}=1.4$. Time-Aware map matching avoids of all these parameters, relying only on data characteristics: the results are confirming the goodness of this approach.  

\subsubsection{Complexity comparison} 
Another improvement achieved with Time-Aware algorithm is a valuable speed up in terms of computation time. 
As stated before, the complexity of Time-Aware map matching is due to the complexity of point-matching process, $O(log |E|)$, 
and the complexity of Dijkstra's algorithm, that is $O(|E| + |V| log |V|)$.
The main difference w.r.t. competitors is on shortest path computation : Time-Aware uses Dijkstra's algorithm only once for 
every two consecutive GPS points. 
IVMM is computing the shortest path between every couple of candidates for every two consecutive GPS points. 
This means a sensibly higher number of steps to find the solution
w.r.t. 
Time-Aware map matching. Tang-Zhu-Xiao algorithm has a similar behavior: 
in order to compute the most probable candidates for two GPS points, the shortest path between every couple
is candidated is computed.  

\section{Experiments on large datasets} \label{large-dataset}
\begin{figure*}[htbp]
\centering
\begin{minipage}[b]{0.45\linewidth}
\centering
\includegraphics[width=90mm]{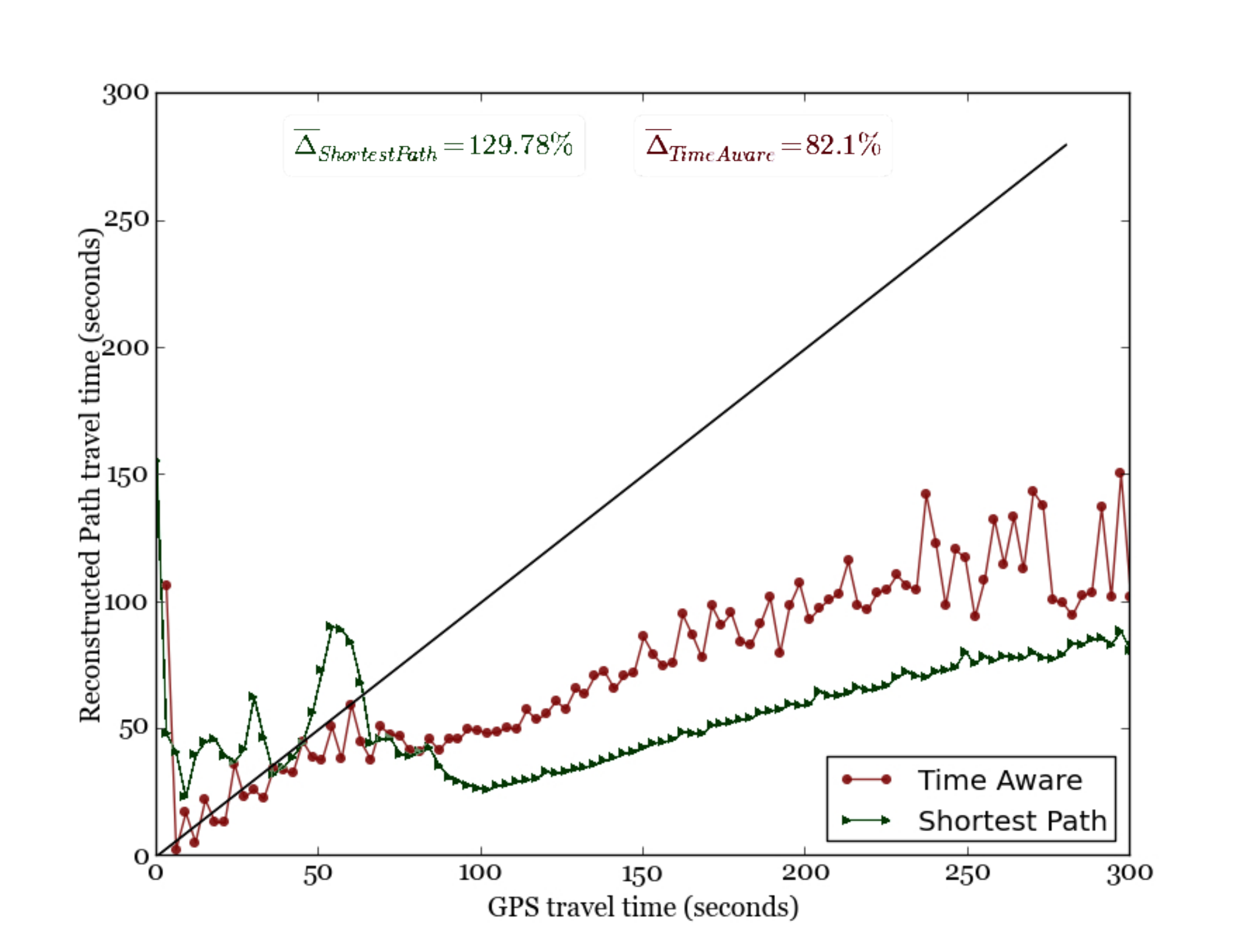}

\end{minipage}
\quad
\begin{minipage}[b]{0.45\linewidth}
  
\centering
\includegraphics[width=90mm]{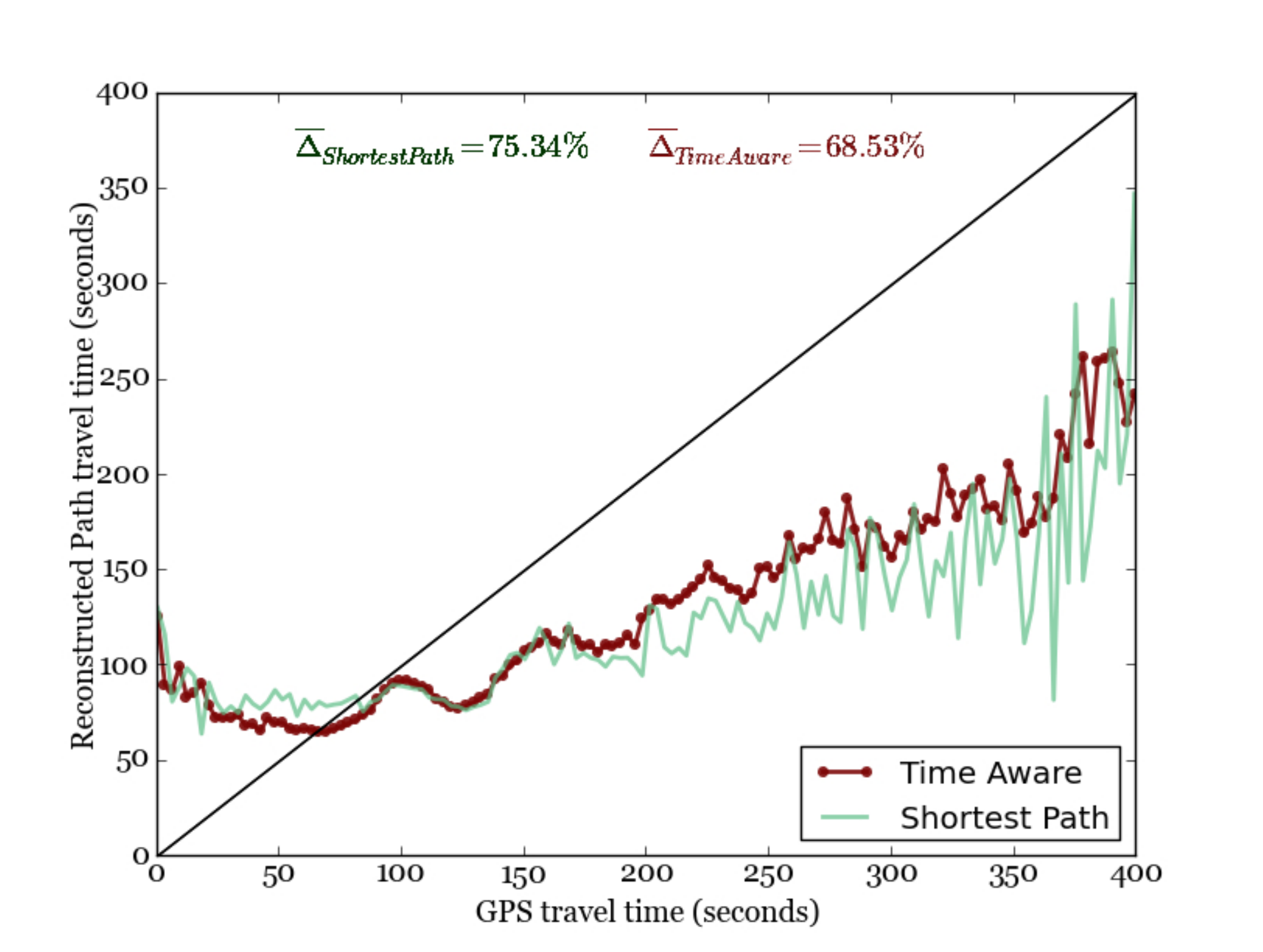}

\end{minipage}
\caption{Average travel times of reconstructed path according to the GPS travel time of the original points, for both Time-Aware and Shortest Path approach applied to Octopisa (left)  and San Francisco Cabs (right) dataset. $\overline{\Delta}$ indicates the average difference between Path and
GPS travel times.}\label{time_difference_Compare}
\end{figure*}

It is not possible to directly evaluate the accuracy of our Time-Aware map matching on a real dataset, since the correct traversed roads are not reported.
However, we performed some tests on two different datasets (see tab. \ref{dataset}) to highlight the improvement of 
our algorithm. The first test we made is a comparison between the travel time of different reconstructed paths according to the used heuristic.
Then, we introduced the middle-point test, that is a coherence test; details for this test 
are provided in the next subsection. 
\begin{table}[bh]
\begin{tabular}{|l|p{2cm}|p{2cm}|}
\hline
\textbf{VMP address} & \textbf{Time Aware}  & \textbf{Shortest Path} \\ \hline
via Aeroporto        & \textbf{0.4604}                          & 0.4602                             \\ \hline
via Cascine          & \textbf{0.3376}                          & 0.3178                             \\ \hline
via di Cisanello     & \textbf{0.5880}                          & 0.5758                             \\ \hline
via Tosco Romagnola  & 0.4511                          & 0.4527                             \\ \hline
via Brennero         & 0.4770                          & 0.4823                             \\ \hline
via San Jacopo       & \textbf{0.6680}                          & 0.6536                             \\ \hline
via Pietrasantina    & 0.6070                          & 0.6083                             \\ \hline
via Emilia           & \textbf{0.7050}                          & 0.6868                             \\ \hline
via Pisano           & \textbf{0.5615}                          & 0.5510                             \\ \hline
Average correlation & \textbf{0.539}                           & 0.532                      \\ \hline

\end{tabular}
\caption{Correlation between VMP data and map matched data}
\label{pmv-corr}
\end{table}
Thanks to the Pisa's VMP dataset (Variable Message Panel) used in \cite{pappalardo2013understanding}, we developed a further test on Octopisa
dataset; we evaluated the coherence of map-matched trajectories w.r.t.data from traffic monitors by using VMPs data. The observed period is the same 
for both OctoPisa and VMP datasets,hence we can provide a reliable
test by checking the correlation among them. Results are showed on table \ref{pmv-corr}.
As showed, Time-Aware approach is slightly outperforming Shortest Path approach. More specifically, the correlation between cars passing by counted by VMPs and
trajectories passing by the VMP road segment is higher if we use our Time-Aware heuristic instead of Shortest Path. In this test we chose Shortest Path as a competitor, 
since it is performing way better than IVMM and Efficient
matching. Although the improvement reported on this test is not high in absolute terms, in the next subsection we will depict how this small difference
is not equally distributed, so enforcing the goodness of our approach: for a test application such as leveraging traffic flows on city access points, the use of Shortest Path 
heuristic lead to a less precise estimation, expecially
for some particular roads.\\

\begin{table}[htpb]
\begin{tabular}{|p{4cm}|c|c|}
\hline
\textbf{Dataset }& OctoPisa & SF cabs \\ \hline
\textbf{N. of trajectories}                                & 1,382,892  & 91,244\\ \hline
\textbf{N. of GPS points}                                   & 19,536,742 & 11,120,908\\ \hline
\textbf{N. of users}                                   & 38,259 & 500 \\ \hline
\textbf{Avg. sampling rate}                             & 94.376 s  & 58.45 s\\ \hline
\textbf{Avg. distance between consecutive points}       & 1.538 km & 0.587 km \\ \hline
\textbf{Avg. point-nearest segment distance}       & 8.65 m  & 16.87 m \\ \hline
\textbf{Std. dev. point-nearest segment \newline distance} & 22.65 m  & 33.72 m\\ \hline
\end{tabular}
\caption{Datasets properties}
\label{dataset}
\end{table}

\begin{figure*}[tb]
\centering
\begin{minipage}[b]{.22\linewidth}
\centering

\includegraphics[width=45mm]{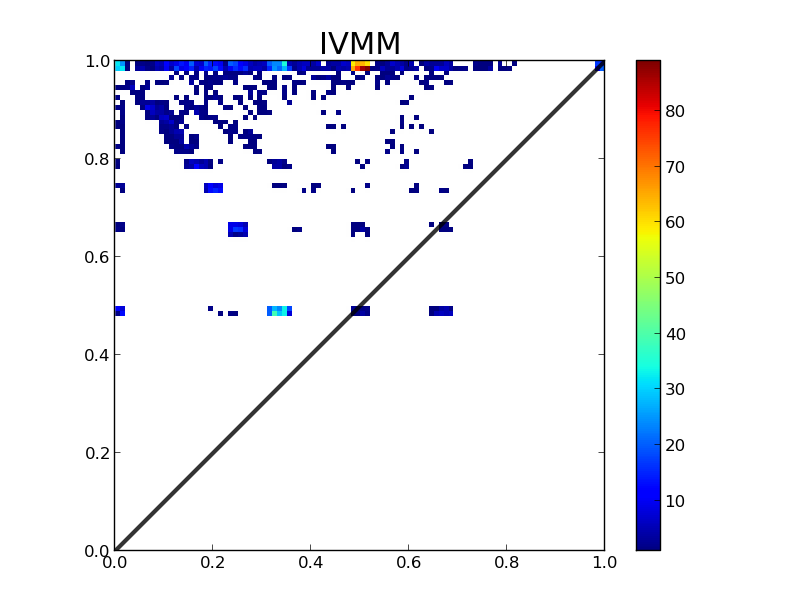}

\end{minipage}
\quad
\begin{minipage}[b]{.22\linewidth}

\includegraphics[width=45mm]{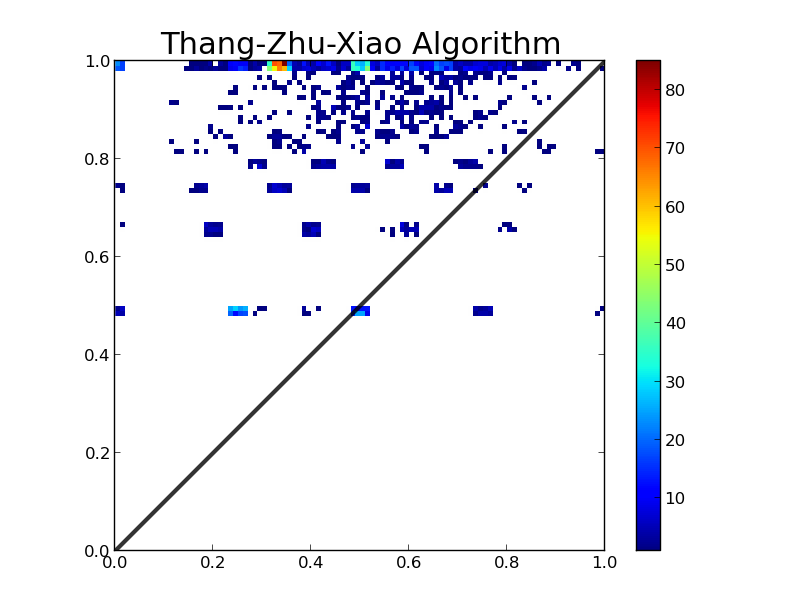}
\end{minipage}
\quad
\begin{minipage}[b]{.22\linewidth}

\includegraphics[width=45mm]{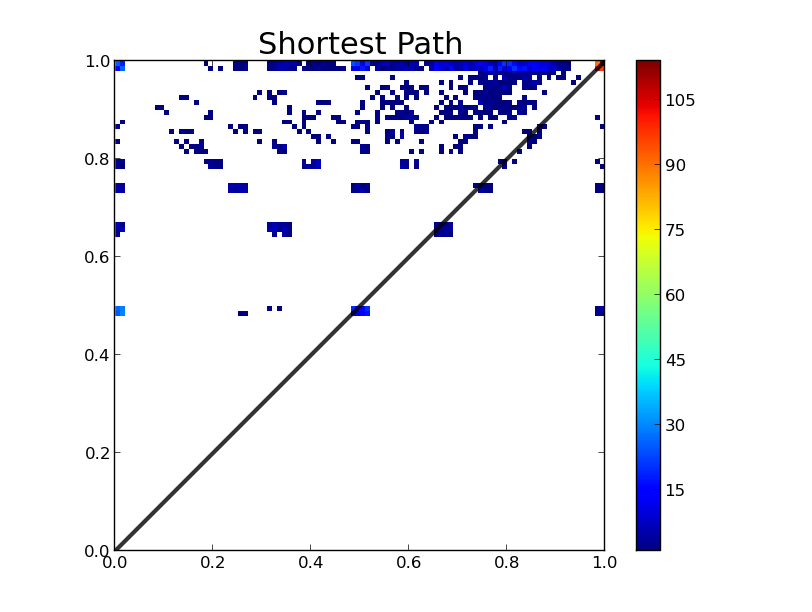}
\end{minipage}
\quad
\begin{minipage}[b]{.22\linewidth}

\includegraphics[width=45mm]{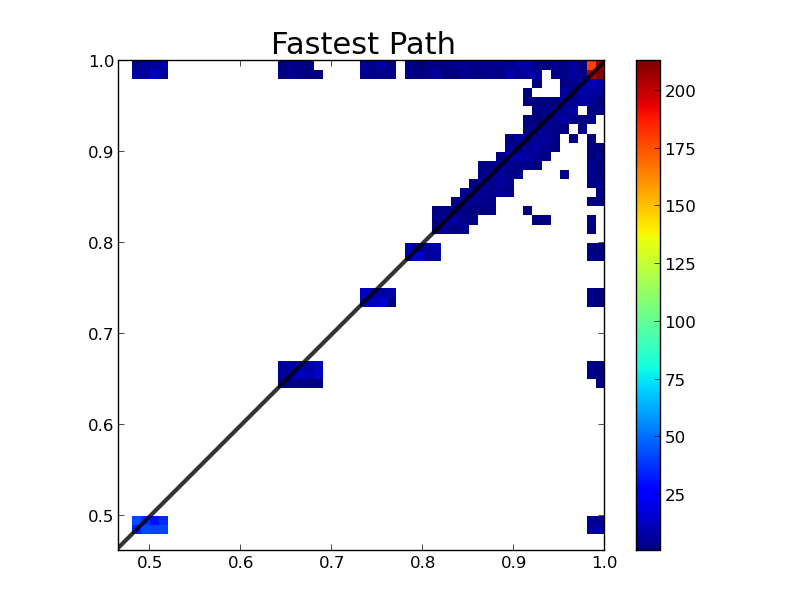}
\end{minipage}

\begin{minipage}[b]{.22\linewidth}
\centering

\includegraphics[width=45mm]{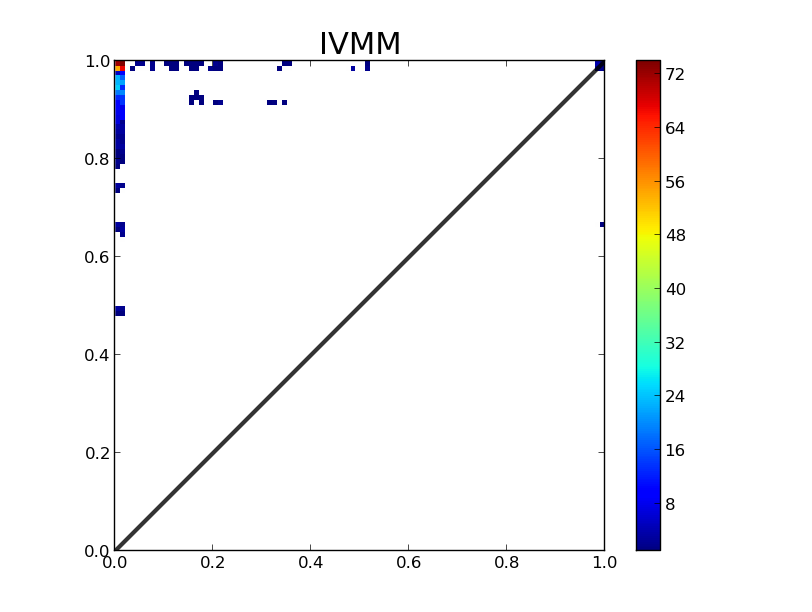}

\end{minipage}
\quad
\begin{minipage}[b]{.22\linewidth}

\includegraphics[width=45mm]{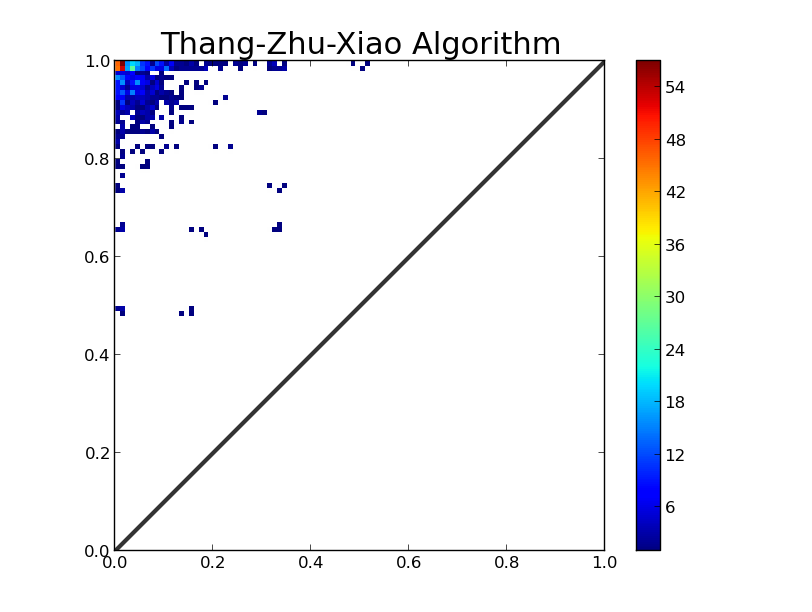}
\end{minipage}
\quad
\begin{minipage}[b]{.22\linewidth}

\includegraphics[width=45mm]{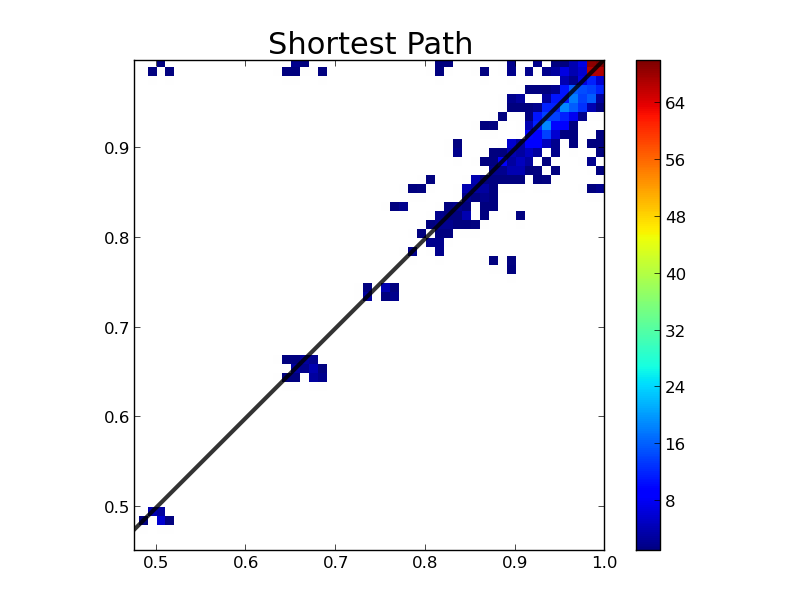}
\end{minipage}
\quad
\begin{minipage}[b]{.22\linewidth}

\includegraphics[width=45mm]{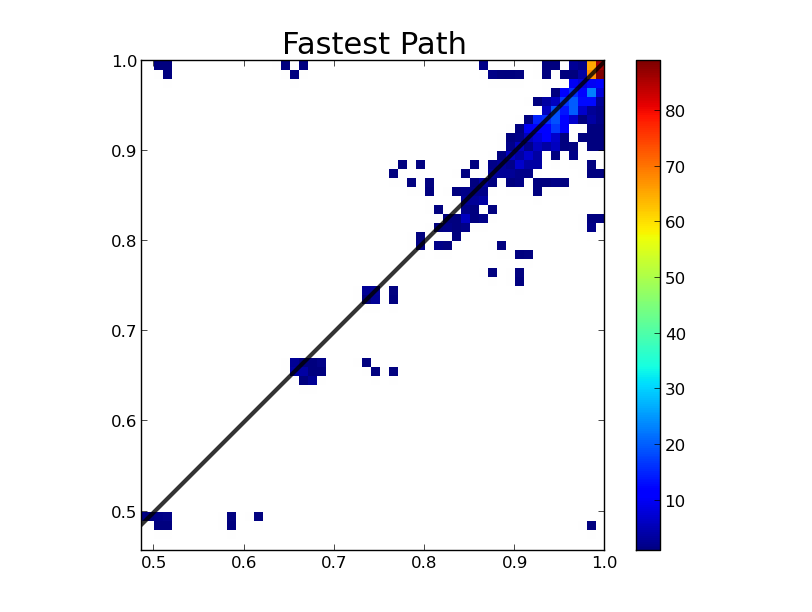}
\end{minipage}

\caption{Comparison between the accuracy of Time-Aware and the other map matching algorithms on Middle-point test for Octopisa (top) and San Francisco cabs (bottom) datasets. 
The line represents the equal case, where the performances of the two approaches are the same. For all the points above the line, performances of Time-Aware approach are better.}
\label{middle-point-plot}
\end{figure*}

%
%

\subsection{Datasets}
As introduced before, we applied our Time-Aware algorithm on two large dataset. 
Octopisa is a database composed by almost 20 million of GPS points recorded for insurance purposes by devices installed on 40K
different cars. The average sampling rate is 90 seconds (see tab. \ref{dataset}). In Fig. \ref {dataset-img} there is a sample of 
this dataset: even though GPS is the most precise way to get geospatial data, the average error is 8 meters with a standard deviation of 22 meters, 
so introducing a lot of noise and making the map-matching task harder. 

We also used a public dataset (\cite{epfl-mobility-2009-02-24}). 
This dataset is composed by GPS traces of 500 cabs traveling in the San Francisco Area for a period of 30 days. 
The distribution of GPS sampling rate is shown
in Fig. \ref{sampling_Distribution}: while the sampling rate for San Francisco cabs trajectories seems to follow a Normal distribution, Octopisa
has a slightly less uniform distribution, probably due to the GPS collecting policies adopted by the provider. 

\subsection{Temporal alignment}
In Fig. \ref{time_difference_Compare} the correlation between GPS travel time and map-matched path travel time is highlighted, for both Time-Aware and Shortest Path approach. 
It is clear how our method yields values closer to the real traversal times, therefore providing a better solution w.r.t. to Shortest Path.
Using the Time-Aware approach, the average difference between real time and path time is almost 50\% less than Shortest Path approach. 
\subsection{Middle-point Test}
To further validate our Time-Aware heuristic, we provide an accuracy test made on some real datasets. For each trajectory we hide the middle point 
of every consecutive GPS points triplet and we repeated the map matching, thus counting
the number of hidden points correctly matched. We used this middle-point test to assess the coherence of the different matching algorithms
 w.r.t. the input trajectory. From table \ref{middle-point-accuracy} is evident
how the Time-Aware heuristic is better on reconstructing a trajectory with half of the original points. Figure \ref{middle-point-plot} 
shows the comparison between our Time-Aware algorithm and the competitors with a density map: the points
above the line represent the trajectories where the performance of Time-Aware is better than competitors.


As depicted in Fig. \ref{middle-point-plot}, Time-Aware algorithm is still overperfoming the results of competitors. Each density plot represent the comparison between Time-Aware 
algorithm and a competitor, conducted on both datasets. The area above the line represents the cases where Time-Aware algorithm has a better accuracy w.r.t to the compared method. 
However, it is worth to point out two differences w.r.t the results obtained on the OctoPisa dataset.
 IVMM and Efficient-match are perfoming worse, this is probably due to their high dependence on parameters, as delineated in the previous section.
 Conversely, shortest and fastest path heuristics obtained better results. This was expected, since all the GPS trajectories have been generated by taxi drivers,
 who have better knowledge about the road network than any other kind of driver. It is worth to notice the different and less accurate scenario on which we performed the test: travel times for San Francisco road network are estimated
 from speed limits, since GPS from taxi cabs do not record instant speed. This makes the application of speed estimation (\cite{cintia2013gravity}) impossible. Despite this lack of information, the Time Aware heuristics still
 confirms its reliability.

\begin{table}
     \begin{tabular}{|p{3cm}|p{1.5cm}|p{2cm}|}
    \hline
    \textbf{Algorithm}                     & \textbf{Octopisa} & \textbf{San Francisco cabs} \\ \hline
    IVMM                          & 0.345  & 0.13  \\\hline
    Tang-Zhu-Xiao algorithm              & 0.41   &  0.14 \\ \hline
    Gravity Model + shortest path & 0.515  &  0.94\\ \hline
    Gravity Model + fastest path & 0.509   & 0.95\\ \hline
    Time Aware                    & \textbf{0.669}  & \textbf{0.95}  \\ \hline
    \end{tabular}
    
    \caption{Average global accuracy on middle-point test}
    \label{middle-point-accuracy}
\end{table}
\begin{figure*}[hb]
\centering

\makebox[\textwidth][c]{
\begin{tabular}{c p{1cm} c}

\includegraphics[width=75mm]{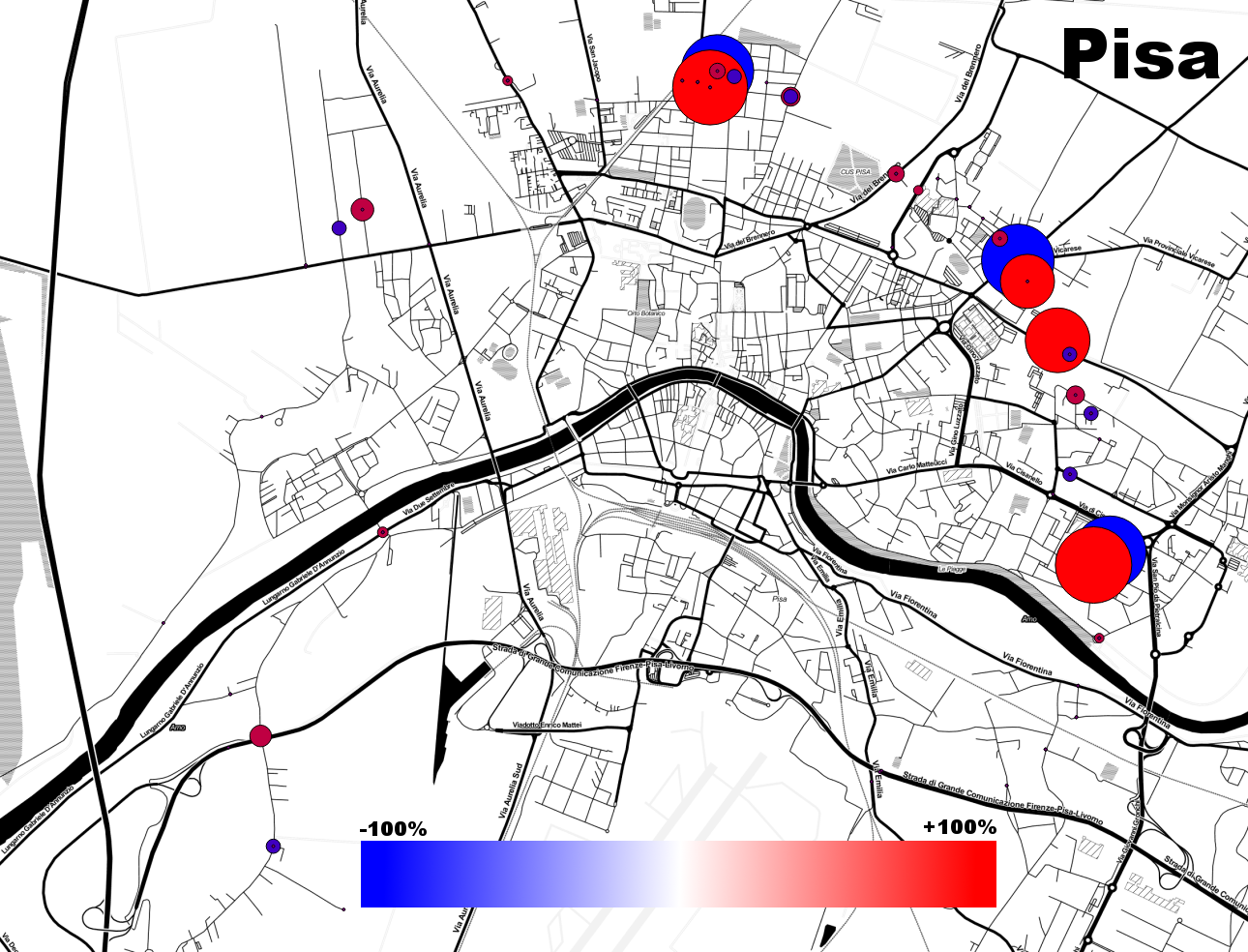} & &
\includegraphics[width=75mm]{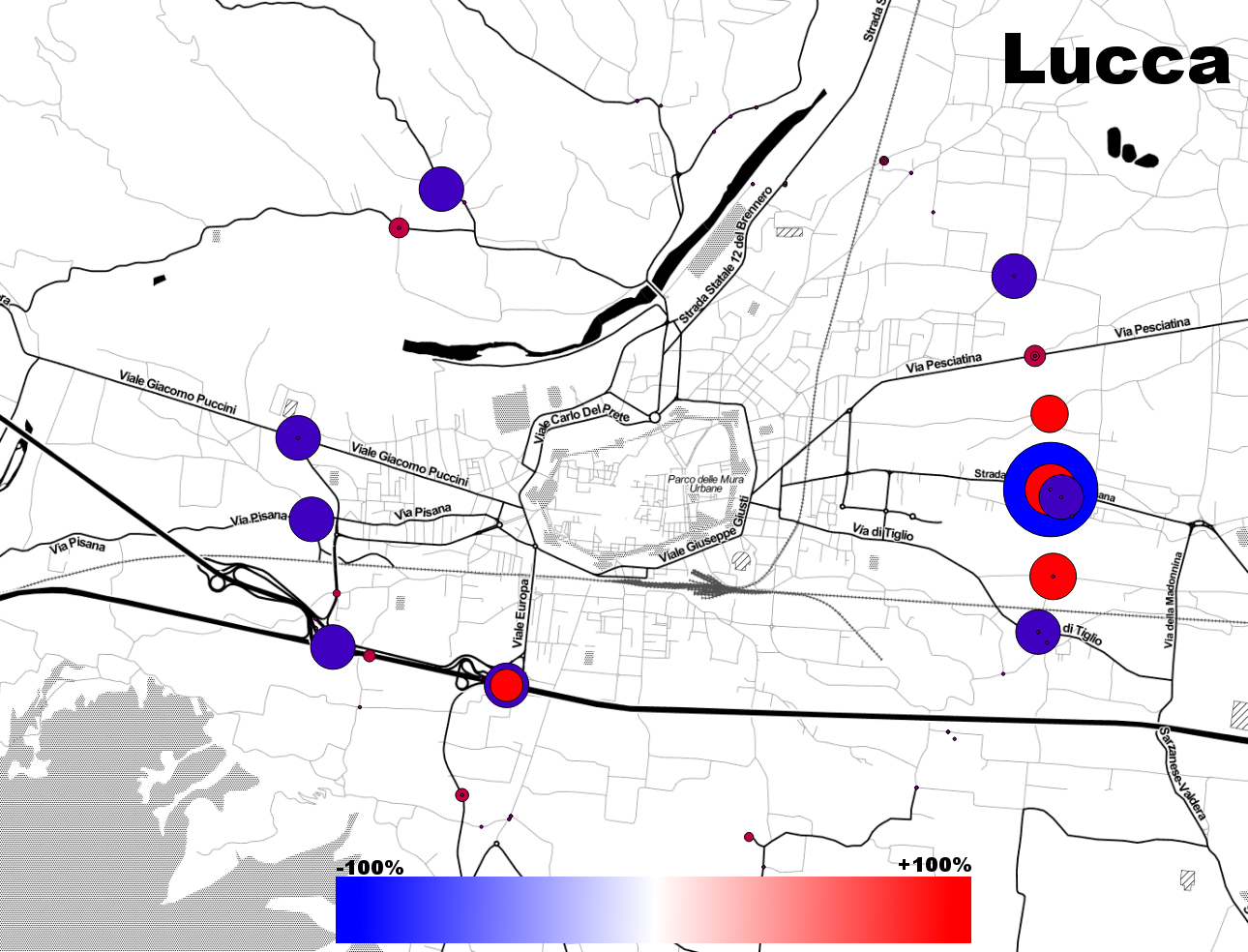}  \\ 
\end{tabular}

}

\caption{Using time-aware map-matched data: this plot highlight traffic flows differences between shortest path and time-aware estimation for the access points 
of Pisa and Lucca. Points size and color represent the higher(road) or lower(blue) traffic according to the application of Shorthest Path approach w.r.t
Time-Aware. By comparing this result with the domain knowledge, Shortest Path tends to overestimate the traffic of secondary roads hence underestimanting the main ones.}
\label{pisa_access_point}
\end{figure*}

\begin{figure*}[htbp]
\begin{minipage}[t]{.45\linewidth}
\centering
\includegraphics[width=80mm]{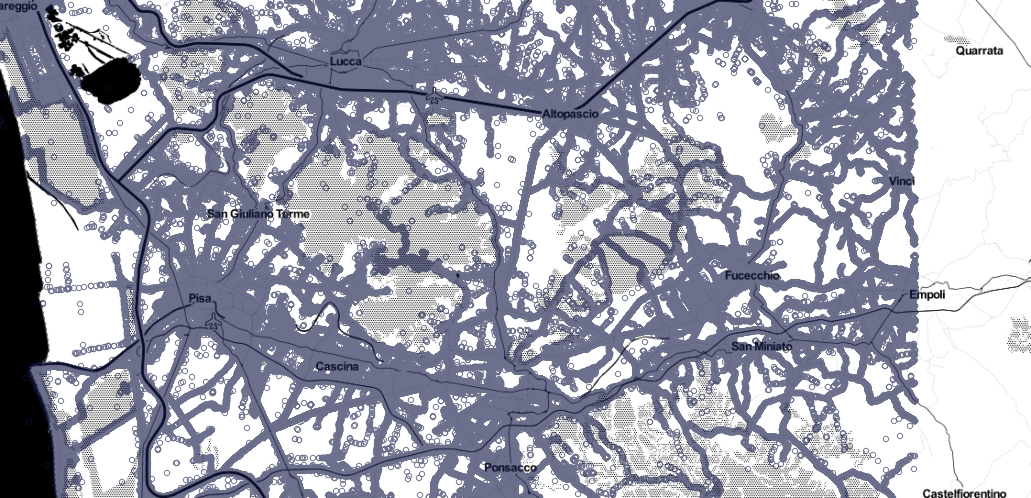}
\caption{A sample of our dataset: raw points can give us only a rough idea of road network usage}
\label{dataset-img}
\end{minipage}
\quad
\begin{minipage}[t]{.45\linewidth}

\centering
\includegraphics[width=80mm]{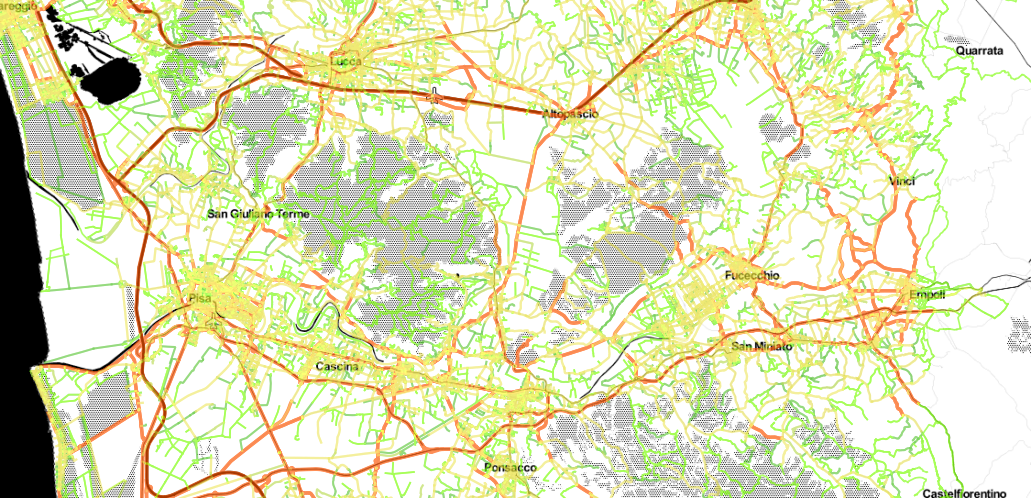}
\caption{An application of our map-matching: for every road segment we can compute the number of vehicle passing by. Red segments are the most used, while green ones are the most free}
\label{road_count}
\end{minipage}
\end{figure*}

\subsection{Applications}

Once validated our method, we used it to map match our OctoPisa dataset (see Tab. \ref{dataset} and sampling rate cumulative distribution on fig.
\ref{sampling_Distribution}). GPS error and sampling rate for this dataset mean a lot of noise: Fig. \ref{dataset-img} gives an idea of the GPS error we avoided with our map-matching algorithm.
We used M-Atlas (\cite{trasarti2010exploring}) to extract trajectories from raw data. Then, we applied Time-Aware map matching to our dataset, adding a reliable semantic layer to the raw data. This gives us the possibility to exploit many useful analysis. 
We propose here two examples of data analysis with map matched data.
In Fig. \ref{road_count} the usage of road network is shown. Segments are colored according to the number of vehicles passing by. 
This introduces a new way for traffic monitoring, since nowadays transport managers are mainly using fixed and costly structures as video devices. 
The availability of GPS and geospatial data is a big chance for transport managers: a deeper and precise view on traffic is now easily achievable. 
Despite VMPs, which are fixed and costly devices,
with this data we can perform traffic analysis in a more flexible way. We can filter the trajectories according to different parameter, like distance traveled, direction (city center, city-to-city, hinterland etc). Then,
we can decide where to put our ''virtual'' traffic monitors to observe the traffic flow in those points. In Fig. \ref{pisa_access_point}  an 
example is provided. 
We selected all the entering trajectories for two cities, 
respectively Pisa and Lucca, in order to draw the access points of the city. We leverage the traffic flows on these access points using both 
Time-Aware and Shortest Path, then we plot differences between fluxes in Fig. \ref{pisa_access_point}.
An expert domain could immediately notice how some minor roads become more important than expected: those segments are part of lots of 
shortest paths, but
their travel time does not fit with GPS travel time. Although the two methods seem to not be really different 
in absolute terms, as we reported with matched trajectories-VMPs correlation comparison (section \ref{large-dataset}), from Fig. \ref{pisa_access_point} it is clear how those differences are not equally distributed. This enforce the importance of our Time-Aware approach, that is 
able to avoid strange matching of minor roads not used in the reality.  \\

\section{Conclusions}
In this paper we proposed a Time-Aware map matching process based on a new approach to the map matching problem. 
We have shown how the state-of-the-art algorithms are
affected by issues, due to their high number of parameters and the adoption of generally false assumption, i.e. the fact that drivers always take the shortest or fastest route to reach a destination. 
Thus, we provided a parameter-free algorithm 
that outperforms the state-of-art competitors, both in terms of accuracy and complexity. The goodness of our approach has been showed 
on three different datasets. 
The future developments of this work go towards the study of methods that better exploit the knowledge we can acquire about a single individual when a long history of her movements are available. Moreover, we plan to test
the proposed algorithm on richer datasets, such as the one from \textit{TagMyDay} (\cite{tagmyday}): a joint project 
between ISTI-CNR and Pisa Transport Manager office based on volunteer recruiting for
mobility data collection. This dataset has a semantic layer added by volunteers whom indicate the destination of their trips in terms 
of activity performed (e.g. going home, shopping, etc). 
Finally, an alternative context to explore with our approach is the map matching of GSM data: since this kind of data has a very poor spatial precision, the use of a Time-Aware heuristic might be the key to extract more meaningful traffic information from GSM mobility data.


\clearpage

\section*{References}

\bibliography{elsarticle-template}{}

\end{document}